\documentclass[%
reprint,
superscriptaddress,
showpacs,preprintnumbers,
bibnotes,
 amsmath,amssymb,
 aps,
 prl,
floatfix,
]{revtex4-1} 

\usepackage{natbib}
\usepackage{graphicx}
\usepackage{dcolumn}
\usepackage{amsmath,amsthm}
\usepackage{amsfonts}
\usepackage{amssymb}
\usepackage{bbold}
\usepackage{times}
\usepackage{mathrsfs}
\usepackage{color}
\usepackage{gensymb}
\usepackage{times}
\usepackage{bm}
\usepackage{stmaryrd}
\usepackage{comment}
\DeclareGraphicsExtensions{.PNG,.jpg,.pdf,.gif,.eps}
\usepackage{amsmath,amssymb,amsthm,amsfonts,eucal}
\usepackage[english]{babel}
\usepackage{mathtools, nccmath}

\newcommand{\subfigimg}[3][,]{%
  \setbox1=\hbox{\includegraphics[#1]{#3}}
  \leavevmode\rlap{\usebox1}
  \rlap{\hspace*{-0.05cm}\raisebox{\dimexpr\ht1-0\baselineskip}{#2}}
  \phantom{\usebox1}
}

\newcommand{\rect}{\text{rect}}
\begin{document}

\title{Peristalsis by  pulses of activity}

\author{N.Gorbushin}
\affiliation{\it  PMMH, CNRS -- UMR 7636, CNRS, ESPCI Paris, PSL Research University, 10 rue Vauquelin, 75005 Paris, France}

\author{L. Truskinovsky}
\affiliation{\it  PMMH, CNRS -- UMR 7636, CNRS, ESPCI Paris, PSL Research University, 10 rue Vauquelin, 75005 Paris, France}

\begin{abstract}
Peristalsis  by  actively generated  waves of muscle contraction   is one of the most fundamental ways of producing motion  in living systems.  We show that peristalsis can be  modeled  by a train of  rectangular-shaped solitary waves of localized activity propagating through otherwise passive matter.  Our analysis is based on  the  FPU-type   discrete  model    accounting for  active stresses and  we  reveal the existence in this problem of a  critical regime which  we argue to be  physiologically advantageous.  
\end{abstract}

\date{\today}

\maketitle
Peristalsis is  a series of actively generated wave-like muscle contractions and relaxations  which propagate along the  body of an organism. Smooth muscle tissues develop  such  contractions to produce a  peristaltic wave in the digestive tract \cite{sinnott2017peristaltic, brandstaeter2019mechanics}. Crawling by  peristalsis enables  animals  like   snails, earthworms, slugs, and terrestrial planarians, to  advance  in narrow spaces
~\cite{gray1938studies, chapman1958hydrostatic, jones1978observations,alexander1982locomotion, quillin1999kinematic,tyrakowski2012discrete}, moreover,   based on geometrical  symmetries only,  peristaltic waves were shown to be an optimal motility  strategy in such systems \cite{agostinelli2018peristaltic}. 

In this Letter we develop a prototypical model of a  peristalsis in a segmented limbless organism.  We assume that it crawls along a flat surface by extending its forward end and then bringing up its rear end. To achieve this goal  the organism generates a solitary  wave
which travels from the front to the rear. The  space-time distribution of activity in such living systems is known to be highly adaptive ~\cite{boyle2012gait} and the mechanism   of this adaptability has recently become a  subject of great interest in robotics~\cite{daltorio2013efficient,boyle2012adaptive}. 

Peristaltic waves are also of general physical interest as elementary nonlinear excitations  of  active matter. Propagating  active pulses reminiscent of  peristaltic waves are ubiquitous in nature from  shimmering  in  honeybees \cite{kastberger2014speeding} to Mexican waves on  stadiums~\cite{cartwright2006mexican}. Comparable phenomena  in the form of propagating activity bands are also observed in  flocking  colonies of swarming robots and other similar systems~\cite{solon2015pattern,ngamsaad2018propagating}. Some of these behaviors   can be  quantified  using  models  of excitable media~\cite{dudchenko2012self} or models involving some kind  of globally synchronized CPG (central pattern generator) \cite{ijspeert2008central}. However, such models have been questioned in  other cases  clearly  dominated by     mechanical sensory feedback  and neuromechanical proprioception \cite{paoletti2014proprioceptive}. Given the distinctly mechanical  functionality of  peristalsis,  we  forgo the   reaction-diffusion framework~\cite{miller2020gait} and neglect the possible role of CPG, and assume instead that physical forces not only drive the associated localized waves of activity but also secure the  signaling pathways  controlling, for instance,  the necessary internal delays. 

As a toy model, capturing only the main effects, we consider a  mass-spring chain capable of  generating  active stresses. It is implied that behind such  activity   is an endogenous  machinery   of the type involved in muscle  tetanization and  we assume that the associated energy flow through the system can be adequately represented by a  non-constitutive component of stress. We show that the ensuing, apparently purely  mechanical,  model  can  generate directional peristaltic locomotion without relying on externally  coordinated  actuators or digital controllers. Our intentionally minimalistic approach  emulates  (and can be  extended towards)  more comprehensive continuum theories  of  active media with internally generated active stresses known for  both  fluids ~\cite{prost2015active,julicher2018hydrodynamic} and solids~\cite{hawkins2014stress,maitra2019oriented,moshe2018geometric,
scheibner2019odd}. While these models directly account for energy consumption and energy dissipation, our approximate model neglects both. 

Passive  solitary waves have been long  employed in the \emph{actuator-driven} soft robotics imitating peristalsis \cite{raney2016stable,nadkarni2014dynamics,deng2020pulse}.   Instead, here we  rely on  \emph{self-driven}  active  solitary waves and show that peristalsis can be  modeled  by a train of  such   waves propagating through otherwise passive matter.  Rather remarkably, our analysis  reveals the existence of a critical  motility regime  in such systems  where active pulses  assume  realistic rectangular shape with variable width. This ensures  broad repertoire of responses  and we argue that so-interpreted  criticality  may be a characteristic feature of  the physiological  peristalsis. 

Bodies of  annelid animals are  usually divided into a
series of metameres, the segments that are fundamentally similar in muscular structure and functionality \cite{tanaka2012mechanics,quillin1999kinematic}. To model such organisms we first  neglect  friction~\cite{kandhari2019turning, keller1983crawling, wadepuhl1989computer}, and represent them schematically as  a chain of springs connected in series.  The dynamics  of such system is described by the FPU  equations \cite{keller1983crawling}
\begin{equation}
\rho a^2\frac{\partial^2\varepsilon_n}{\partial t^2}=\sigma(\varepsilon_{n+1})+\sigma(\varepsilon_{n-1})-2\sigma(\varepsilon_{n}),
\label{eq:EquationsOfMotion_displ}
\end{equation}
where  $\varepsilon_n(t)=(u_{n+1}(t)-u_n(t))/a$ is the  strain in a spring whose ends  undergo displacements  $u_n(t)$and  $a$ is the equilibrium length of a spring. The inertial term,  allowing the system to overcome the discreteness-induced  trapping, proved to be  important  in ultra-soft robotics~\cite{fang2015phase,jiang2017optimal,tanaka2012mechanics,deng2020pulse}. In physiological setting the  apparent mass density  $\rho$ can be   viewed as  a  parameter  introducing     an activity-related   time delay in the  response of stretch receptors \cite{badoual2002bidirectional, serra2012mechanical}. 
\begin{figure}
\hspace{-0.1cm}
\subfigimg[width=0.235\textwidth,height=3cm]{(a)}{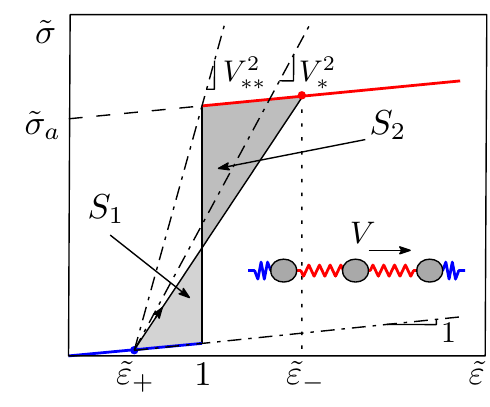}  
\subfigimg[width=0.235\textwidth,height=3cm]{(b)}{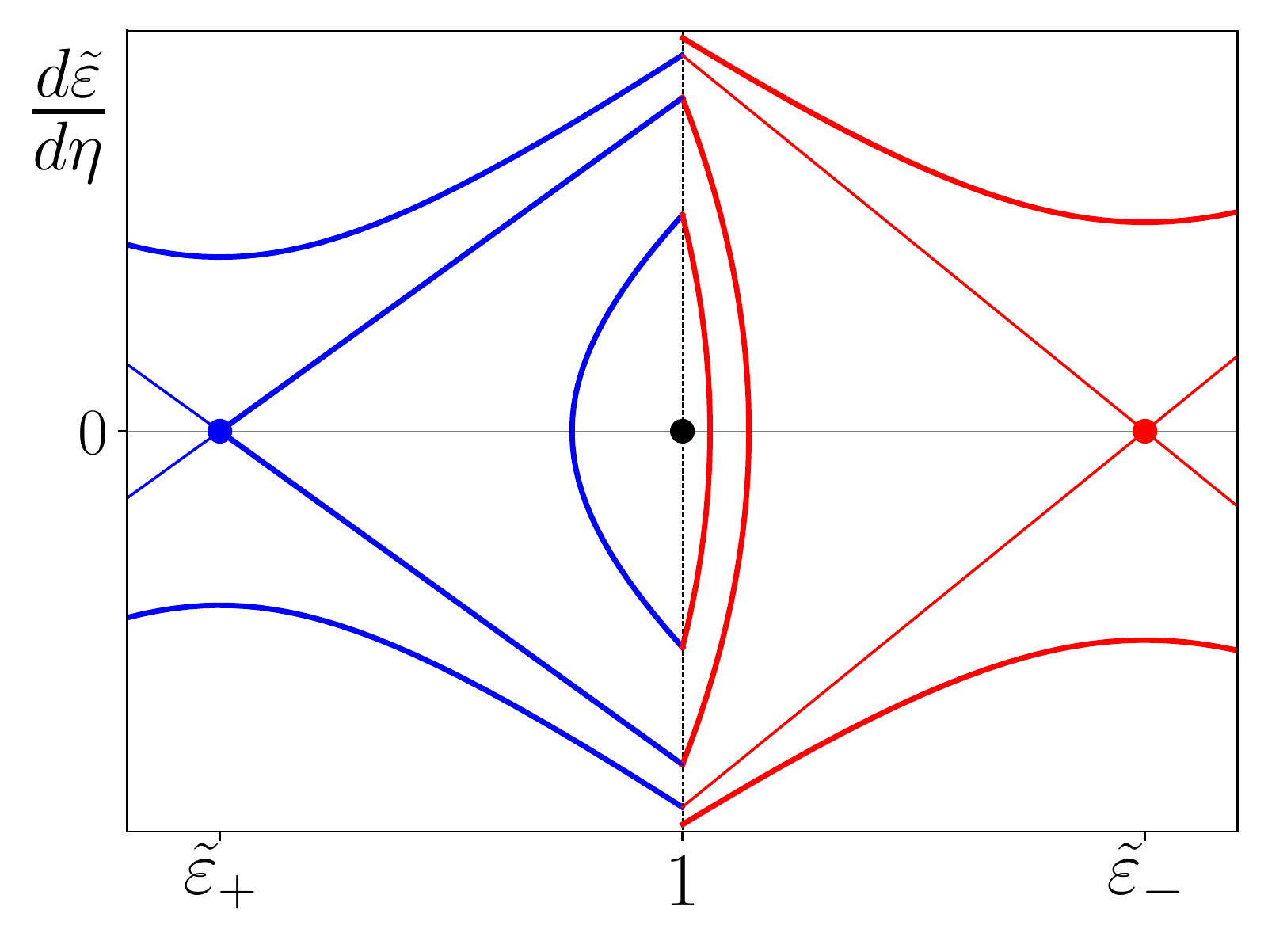}  
\caption{(a) Stress-strain relation  with passive (blue) and active (red) branches. Kink,  connecting $\tilde\varepsilon_{\pm}$ at  the critical value of velocity $V=V_*$ when   $S_2=S_1$,  is dissipation  free. (b) Phase portrait of the  continuum system \eqref{eq:strain_equation_soliton_ros} at $1<V<V_*$. Stretching pulse  corresponds to the homoclinic trajectory starting and ending at $\tilde\varepsilon_+$;  contraction pulse, which exists at $ V_*<V<V_{**}$,  is a homoclinic  trajectory   starting and ending at $\tilde\varepsilon_-$ (not shown).}
\label{fig:chain}
\end{figure}

We assume that  the constitutive  relation for the stress $\sigma$  has  two branches:  passive and active,  see  Fig.\ref{fig:chain}(a).    For simplicity, the soft elastic response along  the   passive branch  is considered linear  $\sigma=E\varepsilon,$  with  elastic modulus $E$.  To describe  the active branch (analog of  muscle tetanus), we  write
$
\sigma = 
\sigma_a+ E\varepsilon,
$
where  $\sigma_a>0$ is a constant  active stress; in  more detailed models it can be a variable with sigmoidal response  taking  a value  zero in the passive phase~\cite{paoletti2014proprioceptive}. We further assume that switching from passive to active response takes place when  the   'unjamming' threshold strain $\varepsilon_c$ is reached \cite{serra2012mechanical}. We  neglect   the possibility of  hysteresis  and assume that unloading below the  threshold $\varepsilon_c$ brings the system  back  into the passive  state, see  Fig.\ref{fig:chain}(a). To nondimensionalize the system  \eqref{eq:EquationsOfMotion_displ} we normalize length by  the system size  $L$, time  by  $L/c$, where $
c=\sqrt{ E/\rho}$, and stress by  $E$.

It will be convenient to first deal with a  continuum approximation of the   discrete problem.  To this end we  introduce the continuous strain field $\varepsilon(x,t)$,   where $\varepsilon(nh,t)=\varepsilon_{n}(t)$   and assume that  $h=a/L\ll 1$; we will   also use the convenient rescaling  $\tilde u=u/\varepsilon_c,\,\tilde{\varepsilon}=\varepsilon/\varepsilon_c$ and $\tilde{\sigma}=\sigma/\varepsilon_c$. 
If we  now  Pad{\'e}-approximate the nonlocal operator in the right hand side of \eqref{eq:EquationsOfMotion_displ} and leave only the lowest order terms, we obtain  \cite{rosenau1986dynamics, truskinovsky2006quasicontinuum,gomez2012uniform} 
\begin{equation}
\left(1- \frac{ h^2}{12}\frac{\partial^2}{\partial x^2}\right)\frac{\partial^2\tilde\varepsilon}{\partial t^2}=\frac{\partial^2 \tilde \sigma}{\partial x^2}.
\label{eq:problem_diff_quasi_ros}
\end{equation}
To generate  a solitary wave solution of \eqref{eq:problem_diff_quasi_ros} we impose a  traveling wave  anzats $\varepsilon(x,t)=\varepsilon(\eta)$  where $\eta=(x-Vt)/h$ and $V$ is  the  dimensionless velocity of the pulse. 

If we center   the   active  pulse performing local \emph{extension}   at $\eta=0$ and  denote its width by $2d$  we can  write the associated stress  distribution in the form
$
\tilde\sigma(\eta)= \tilde\varepsilon(\eta)+\tilde\sigma_a\rect (\eta/(2d)) ,
$
where $\rect(x)=H(x+1/2)-H(x-1/2)$  and   $H(x)$ is  the Heaviside function. We require that   $\tilde\varepsilon(\eta) \to 0$  as $\eta \to \pm \infty$ and impose at  $\eta=\pm d$  the matching conditions  $\llbracket \tilde\varepsilon \rrbracket=\llbracket d\tilde\varepsilon/d\eta\rrbracket =0$ and  set $\tilde\varepsilon(\pm d)=1$. Then,  integrating    \eqref{eq:problem_diff_quasi_ros}  
  twice and applying the boundary/matching  conditions,  we obtain the equation
\begin{equation}
\left(V^2-1-\frac{V^2}{12}\frac{d^2}{d\eta^2}\right)\tilde{\varepsilon}=\tilde\sigma_a\rect\left(\frac{\eta}{2d}\right),
\label{eq:strain_equation_soliton_ros}
\end{equation}
where   the right hand side 
 implicitly depends on $\tilde\varepsilon$. 
Similar solitary wave solution, describing local   \emph{contraction },  can be obtained if we set  $\tilde\sigma(\eta)=\tilde\varepsilon(\eta)+\tilde\sigma_a(1-\rect(\eta/(2d)))$ and require that  $\tilde\varepsilon\to \lambda$ when $\eta\to\pm\infty$ where $\lambda=\tilde\sigma_a/(V^2-1)$. The  extension  pulses   exist in the range $1<V<V_*$ where $V_*\equiv\sqrt{ \tilde\sigma_a/2+1}$, so they  are supersonic which does not mean that they are fast  given that  the underlying elastic medium is almost  an acoustic vacuum \cite{nesterenko2013dynamics}.  The  contraction pulses exist in the range $V_*<V<V_{**}$ where $V_{**}\equiv\sqrt{\tilde\sigma_a+1} > V_*$.

The phase portrait of the system  \eqref{eq:strain_equation_soliton_ros} at $1<V<V_*$ is shown in Fig.~\ref{fig:chain}(b). Two non-degenerate saddle points at $\tilde{\varepsilon}_{\pm}$   lie on the same 
Rayleigh line $ V^2=(\tilde\sigma(\tilde\varepsilon)-\tilde\sigma(\tilde\varepsilon_+))/(\tilde\varepsilon-\tilde\varepsilon_+)$, see Fig.~\ref{fig:chain}(a). Solitary waves describing the  extension pulses, correspond to  homoclinic trajectories starting and ending  at $\tilde{\varepsilon}_{+}$. Periodic trains of such pulses correspond to closed trajectories encircling the degenerate center at $\tilde{\varepsilon}=1$.  As $ V \to V_*$   homoclinic trajectories become heteroclinic and the  solitary waves turn into  kinks; at $ V =V_*$ we have  $S_1=S_2$ in Fig.~\ref{fig:chain}(a) and therefore the associated macroscopic  discontinuity is dissipation free~\cite{truskinovskii1982equilibrium}.  The  structure of  these solutions   is similar to the one    in  flocking models \cite{caussin2014emergent,solon2015pattern}  modulo the fact that  here we omit  the explicit description of inflow and outflow of energy.   Note  that the  contraction pulses, which exist in the complimentary range of parameters  $V_*<V<V_{**}$, correspond to homoclinic trajectories starting and ending at $\tilde\varepsilon_-$.

In view of the piecewise linear nature of our model,  the  homoclinic  solution of  \eqref{eq:strain_equation_soliton_ros}  can be written explicitly 
\begin{equation}
\tilde{\varepsilon}(\eta)=\begin{cases}
\vspace{0.2cm}
  e^{-(\eta-d)/z},\quad \eta>d,\\
\vspace{0.2cm}
\lambda+(1-\lambda)\dfrac{\cosh(\eta/z)}{\cosh(d/z)},\quad -d<\eta<d,\\
  e^{(\eta+d)/z},\quad \eta<-d,
\end{cases}
\label{eq:Solution_quasi_ros}
\end{equation}
where 
$z=V/\sqrt{12(V^2-1)}$ and  $
\tanh (d/z)= -(1-\lambda)^{-1}.
$ Defining the   amplitude  of the pulse  as $A=\max\tilde{\varepsilon}(\eta)-\min \tilde{\varepsilon}(\eta)$,   we find for extension pulses that  $A=\lambda+ 
(1- \lambda)/\cosh(d/z)$. 
The corresponding explicit solution for contraction pulses is presented in \citep{SOM}.

\begin{figure}[!h]
\begin{center}
\subfigimg[width=0.495\linewidth]{(a)}{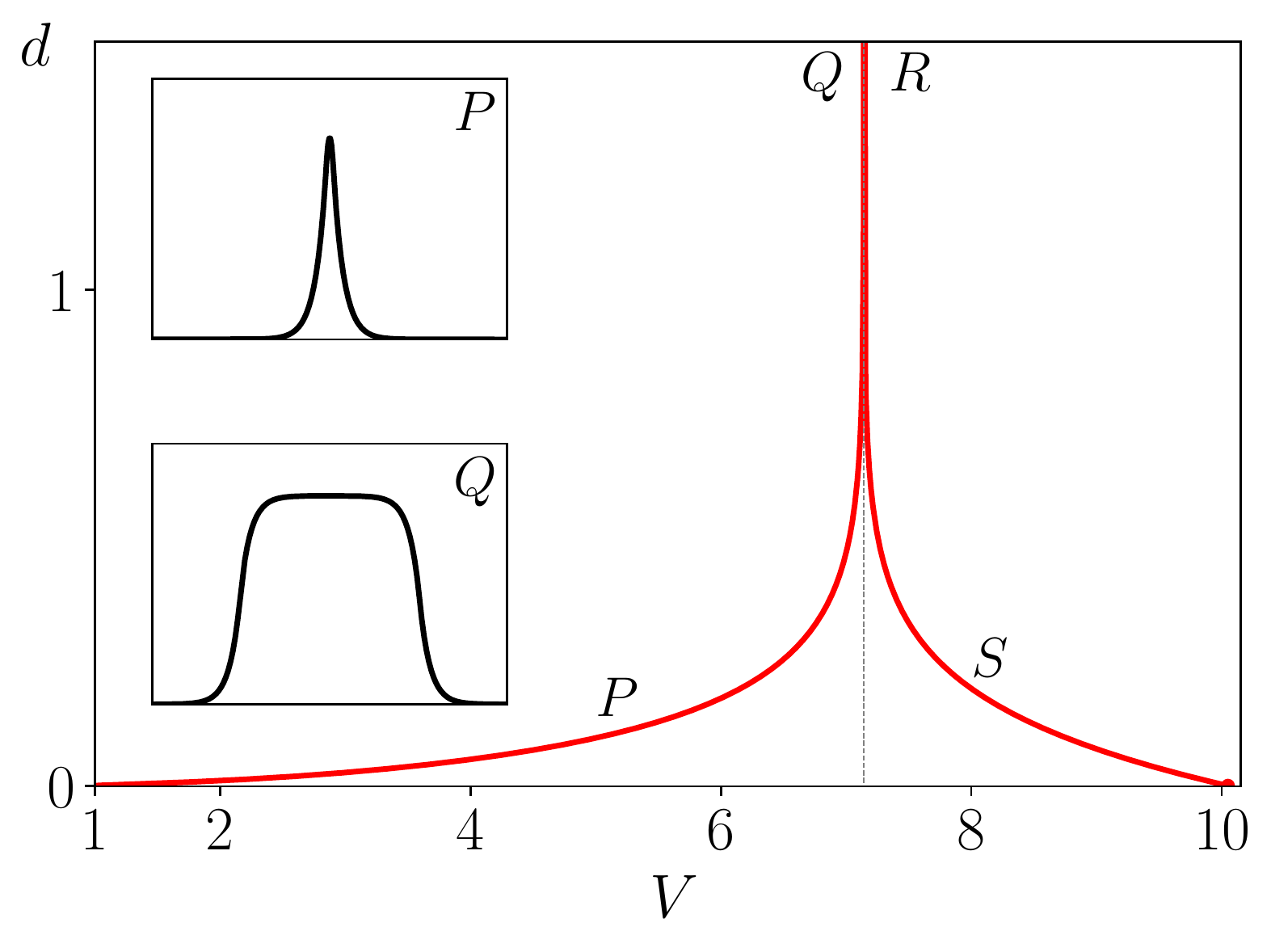}  
\subfigimg[width=0.495\linewidth]{(b)}{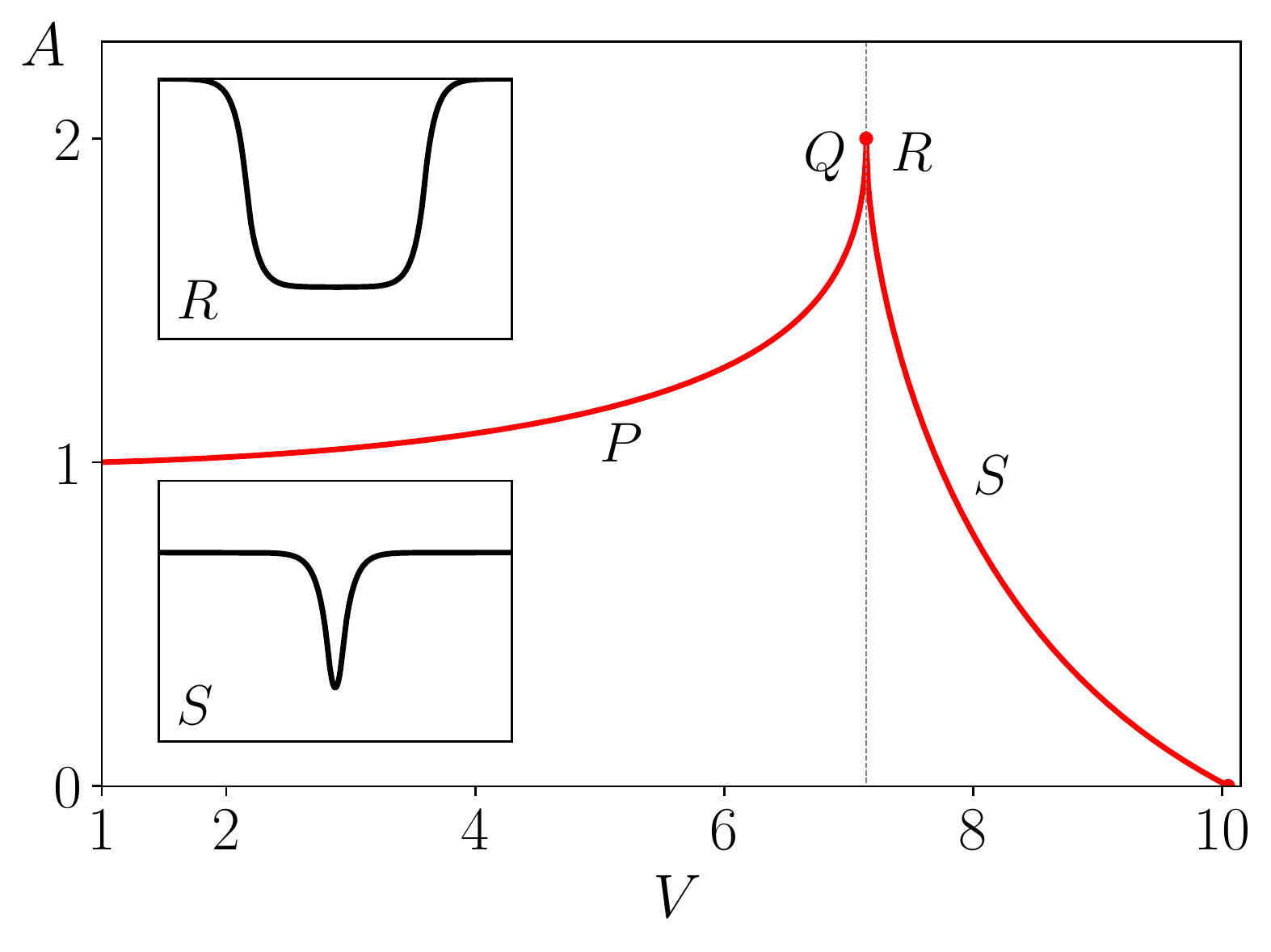}
\end{center}
\caption{Parametric dependence of the structure of  active pulses at $\tilde\sigma_a=100$: (a) the  half-width $d=d(V)$; (b)  the amplitude $A=A(V)$; insets show the strain profiles at: $V=5$ (point $P$), $V=V_*\pm\delta$, where $\delta=10^{-7}$  (points $Q$ and $R$)  and  $V=8$ (point $S$). Vertical  asymptotes mark the location of the critical point  $V=V_*$.}
\label{fig:solution_quasi_ros}
\end{figure}

 The functions $d(V)$ and $A(V)$ for both types of pulses  are  shown  in Fig.~\ref{fig:solution_quasi_ros}(a,b). The two families are separated by the  \emph{critical} value of parameter   $V=V_*$ where the solitary waves take the form of  infinitely separated kinks. At this point the parameter $d$, playing the role of the correlation length,  diverges even though the pulse amplitude remains finite (taking the value $A=2$). In the  limits $V \to 1$ and $V \to V_{**}$ we obtain  sonic waves in passive and active states, respectively; note that the passive limit is singular, see \citep{SOM} for more details. The typical functions  $\tilde\varepsilon(\eta)$ for different values of $V$ are shown  in the insets in Fig.~\ref{fig:solution_quasi_ros}. We emphasize that only  the  near-critical pulses have  a physiologically realistic  rectangular form.

Using the relation $\tilde\varepsilon=-Vd\tilde u/d\eta$,  we can obtain the amplitude of the  displacement increment culminating the passing of a pulse, for instance, in the case of a stretching pulse  $\Delta \tilde u= 2d\lambda$.  As a  rough description of a peristaltic wave train,  represented by a succession of $N$ such pulses, we  can write  $\Delta \tilde u= 2dN\lambda$.  

To construct the actual periodic solution  we  can use the representation 
$
\tilde\sigma(\eta)=\tilde\varepsilon(\eta)+\tilde\sigma_a\sum_{j}\rect(\eta-jD/(2d_p)),
$
where it has been  assumed  that each pulse has a half width  $d_p$ and the whole active  lattice has the period $2D$. The matching conditions  are now  $\llbracket \tilde\varepsilon \rrbracket=\llbracket d\tilde\varepsilon/d\eta\rrbracket =0$ and $\tilde{\tilde\varepsilon}(jD\pm d_p)=1$.  While the whole  solution  can be  again written explicitly, see  \cite{SOM},  here we only present the transcendental equation for the correlation length $d_p$ which is now regularized by the fixed system size $D$: $(1-\lambda)\tanh(d_p/z)= \tanh((d_p-D)/z).$  

\begin{figure}[!h]
  \centering
    \subfigimg[width=0.495\linewidth]{(a)}{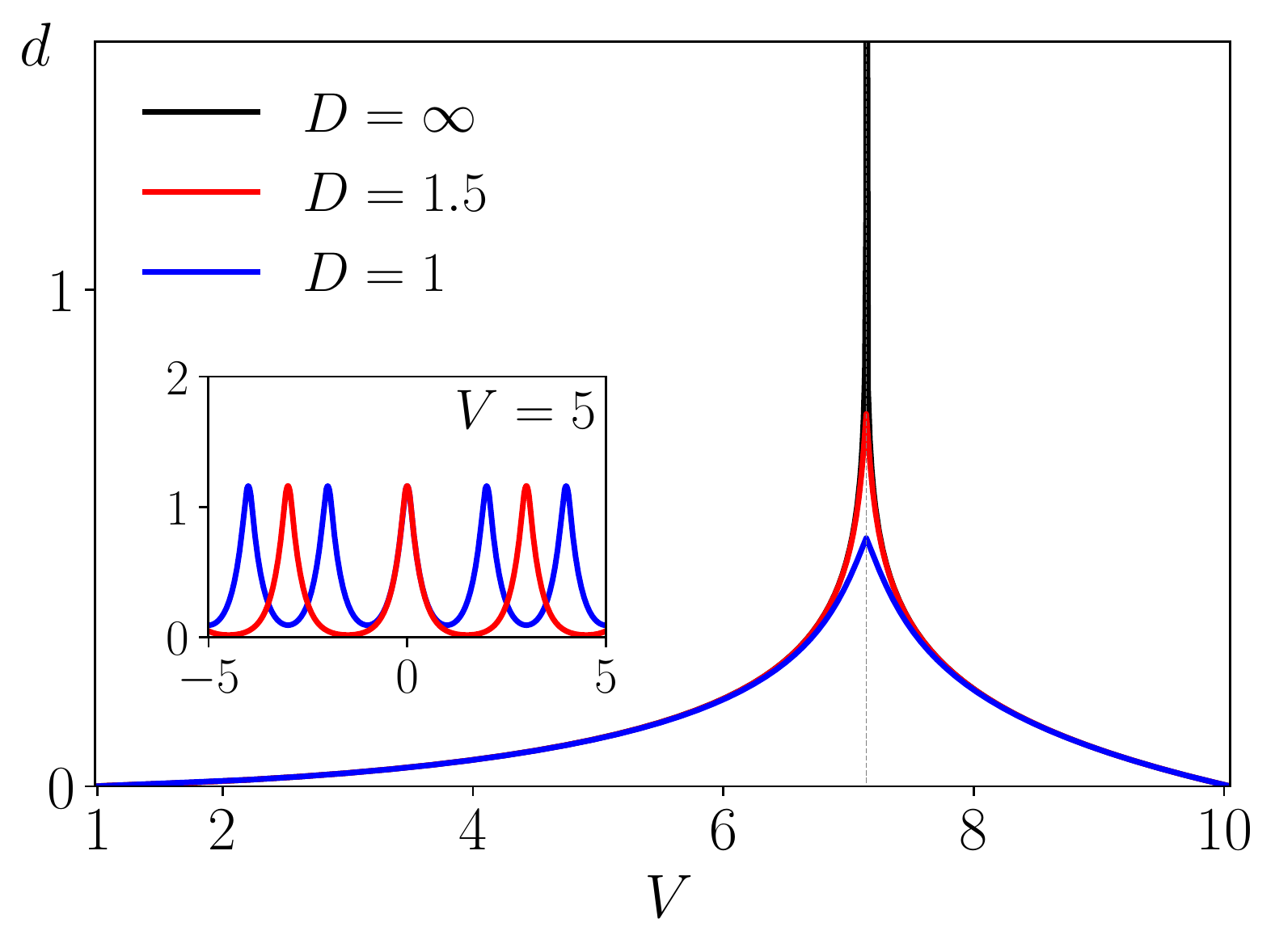}  
    \subfigimg[width=0.495\linewidth]{(b)}{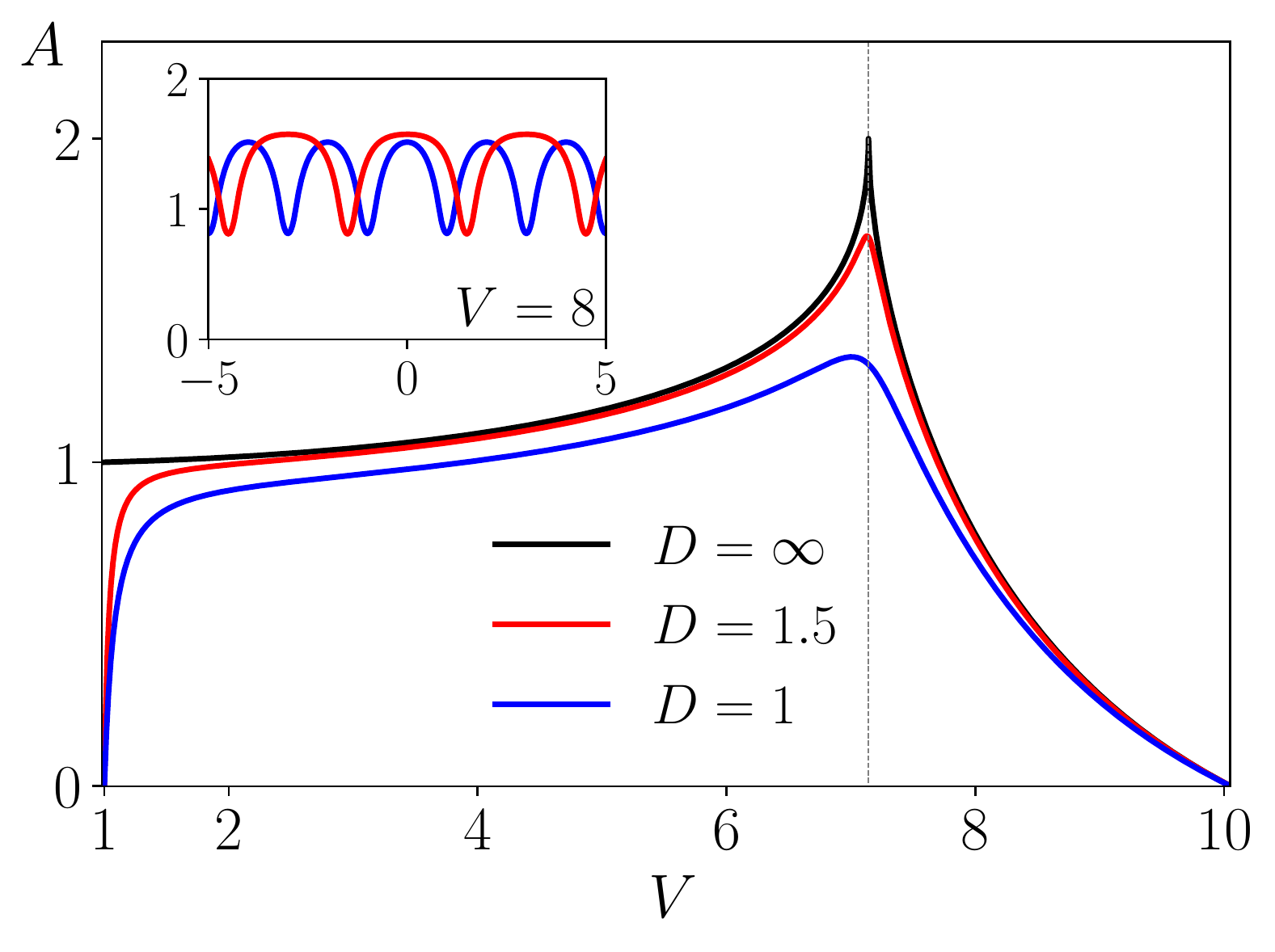}  
  \caption{ Trains of active pulses  at $\tilde\sigma_a=100$: (a)  half-width of a single pulse $d(V)$; (b) its  amplitude $A(V)$.   Vertical asymptotes  correspond to  $V=V_*$; insets illustrate  the typical profiles  $\tilde{\varepsilon}(\eta)$. }
\label{fig:quasi_multiple_d_A_U_ros}
\end{figure}

The obtained  family of solutions is  parameterized  by $V$ and  incorporates both stretching (active) and contraction (passive) pulses. To   distinguish between the two  it is  convenient to re-define the half-width as $d=\min{(d_p,D-d_p)}$ and the amplitude as  $A=|\tilde{\varepsilon}(0)-\tilde{\varepsilon}(D)|$. The resulting functions $d(V)$ and $A(V)$ are shown  in Fig.~\ref{fig:quasi_multiple_d_A_U_ros}.  

\begin{figure}[!h]
  \centering
  \includegraphics[width=\linewidth]{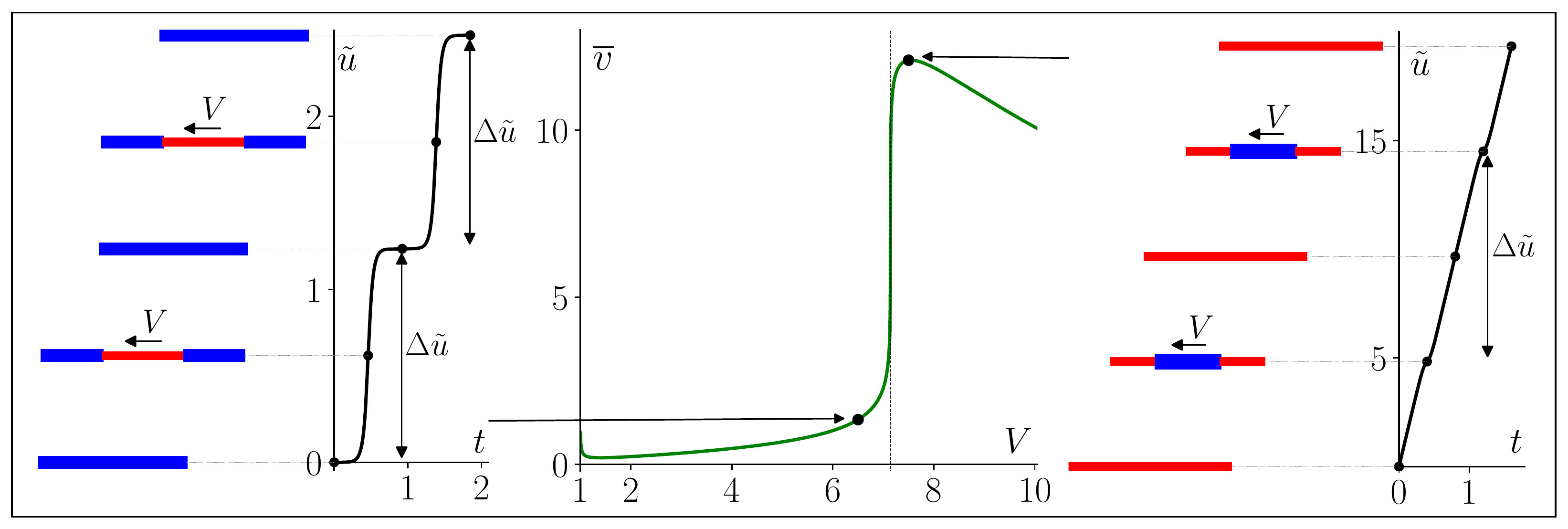}
  \caption{Schematic presentation of peristalsis  by   trains of active pulses at $\sigma_a=100$ and $D=3$. Motility by stretching pulses is shown on the left, by contraction pulses - on the right. The central plot shows dependence of the  average velocity $\overline{v} $ on $V$; vertical asymptote corresponds to the critical point at $V=V_*$.}
\label{fig:peristalsis}
\end{figure}

The obtained periodic solutions  can be used to model peristaltic locomotion.   Suppose   that the organism   generates a periodic  train of \emph{stretching} pulses that propagate rearward with a velocity $V$ so  that  when  a pulse reaches the tail, another one is initiated  at its head with a fixed delay controlled by  the parameter  $D$.  Observations show that when   such  pulse   moves  towards  the tail, the latter fattens and gets  anchored due to local increase of friction. With such an anchor present, the locomotion naturally occurs into the direction opposite to the direction of the pulse. We mimic such motility pattern in Fig.~\ref{fig:peristalsis} (left), where the  pulse is taken from the range  $V<V_*$. 

Suppose  that the motion of the animal is  of stick-slip-type with each advance corresponding to passing of a single pulse producing  the forward displacement  of the head  $\Delta \tilde u= 2d_p\lambda$. Since the next pulse arrives  after the time  $2D/V$, the mean translational velocity of the system  is $\overline{v}=\Delta \tilde uV/(2D)$. Note that  the simultaneous propagation of multiple peristaltic pulses  along the  animal  body is also a possibility which we   do not consider  here. 
 
The  branch of \emph{contraction} pulses  corresponding to   $V>V_*$ also generates a motility  pattern shown  in Fig.~\ref{fig:peristalsis} (right),
In this case the 'fattening' and the resulting anchoring   takes place  around the pulse. Such regimes, driven by  passive pulses in otherwise active medium, are not realistic due to the necessity to maintain active state throughout the whole body of the organism.

The  behavior  of the  function  $\overline{v}(V)$ for both kinds of motility ($V\lessgtr V_*$ ) is shown in Fig.~\ref{fig:peristalsis} (center). We emphasize  that around the critical point   $V=V_*$,  the macroscopic velocity  $\overline{v}$ behaves singularly:  it  varies over a broad range around a single value of the control parameter. The  corresponding individual  pulses take the form of  elongated  rectangles  which is one of the most characteristic features of peristaltic waves.    Since   the  width of these rectangular pulses, and therefore the resulting motility velocity,  can vary  significantly,    being positioned  near such  critical point,  can help an organism to adapt its    responses. The implied anomalous sensitivity to controls  can facilitate  optimal behavior   in complex physiological  conditions and would then be  highly functional.   

Given that  the biological systems exhibiting peristalsis are often  segmented, the question  arises whether our oversimplified  (quasi) continuum model  \eqref{eq:problem_diff_quasi_ros} adequately represents the dynamics of its discrete prototype \eqref{eq:EquationsOfMotion_displ}. To answer this question   we now briefly consider the traveling wave solutions of the original discrete  system.
 
We  maintain the same normalization and use  again  the ansatz  $\tilde\varepsilon_n(t)=\tilde\varepsilon(\eta)$, where $\eta=(nh-Vt)/h$. The discrete strain field  $\tilde\varepsilon(\eta)$  satisfies the equation
 \begin{equation}
 \label{FPU}
V^2\frac{d^2\tilde\varepsilon}{d\eta^2}=\tilde\sigma(\eta+1)+\tilde\sigma(\eta-1)-2\tilde\sigma(\eta),
\end{equation}
with  $\tilde\sigma(\eta)=\tilde\varepsilon(\eta)+\tilde\sigma_aH(-\eta)$.  Solitary wave  can be built directly from kinks which we therefore consider first.

Suppose that a  switching  point  where $\tilde\varepsilon(0)=1$    is  placed  at   $\eta=0$. Application of the Fourier transform to  \eqref{FPU} allows one to obtain an  explicit solution:
\begin{equation}
\label{RR}
\tilde\varepsilon_k(\eta)=\tilde\varepsilon_+-\frac{\tilde\sigma_a}{2\pi}\int_{-\infty}^{\infty}\frac{\omega^2(k)}{(0+ik)L(k)}e^{-ik\eta}\,dk,
\end{equation}
where $L(k)=\omega^2(k)-(kV)^2$, $ \omega^2(k)=4\sin^2\left(k/2\right)$ and  $\tilde\varepsilon_+=\tilde\varepsilon(\infty)$. 
Computing the integral in \eqref{RR}  we obtain the representation of this   solution in the form of infinite series
\begin{equation}
\tilde{\varepsilon}_k(\eta)=\begin{cases}
\vspace{0.2cm}
\tilde{\varepsilon}_++\tilde\sigma_a\sum\limits_{k\in Z^-}\dfrac{\omega^2(k)}{kL'(k)}e^{-ik\eta},\, \eta>0,\\
\tilde{\varepsilon}_--\tilde\sigma_a\sum\limits_{k\in Z^+}\dfrac{\omega^2(k)}{kL'(k)}e^{-ik\eta},\, \eta>0,
\end{cases}
\end{equation}
 where $Z^\pm=\left\{k\,:\,L(k)=0,\,\pm \mathrm{Im}\, k>0\right\}$ and $\tilde\varepsilon_-=\tilde\varepsilon_++\lambda$.  This  solution exists only for the critical value of velocity $V=V_*$, see  \cite{SOM} for details,  which suggests  that   admissible kinks   in this model   must necessarily satisfy the dynamic Maxwell condition.
\begin{figure}[!h]
\centering
    \subfigimg[width=0.495\linewidth]{(a)}{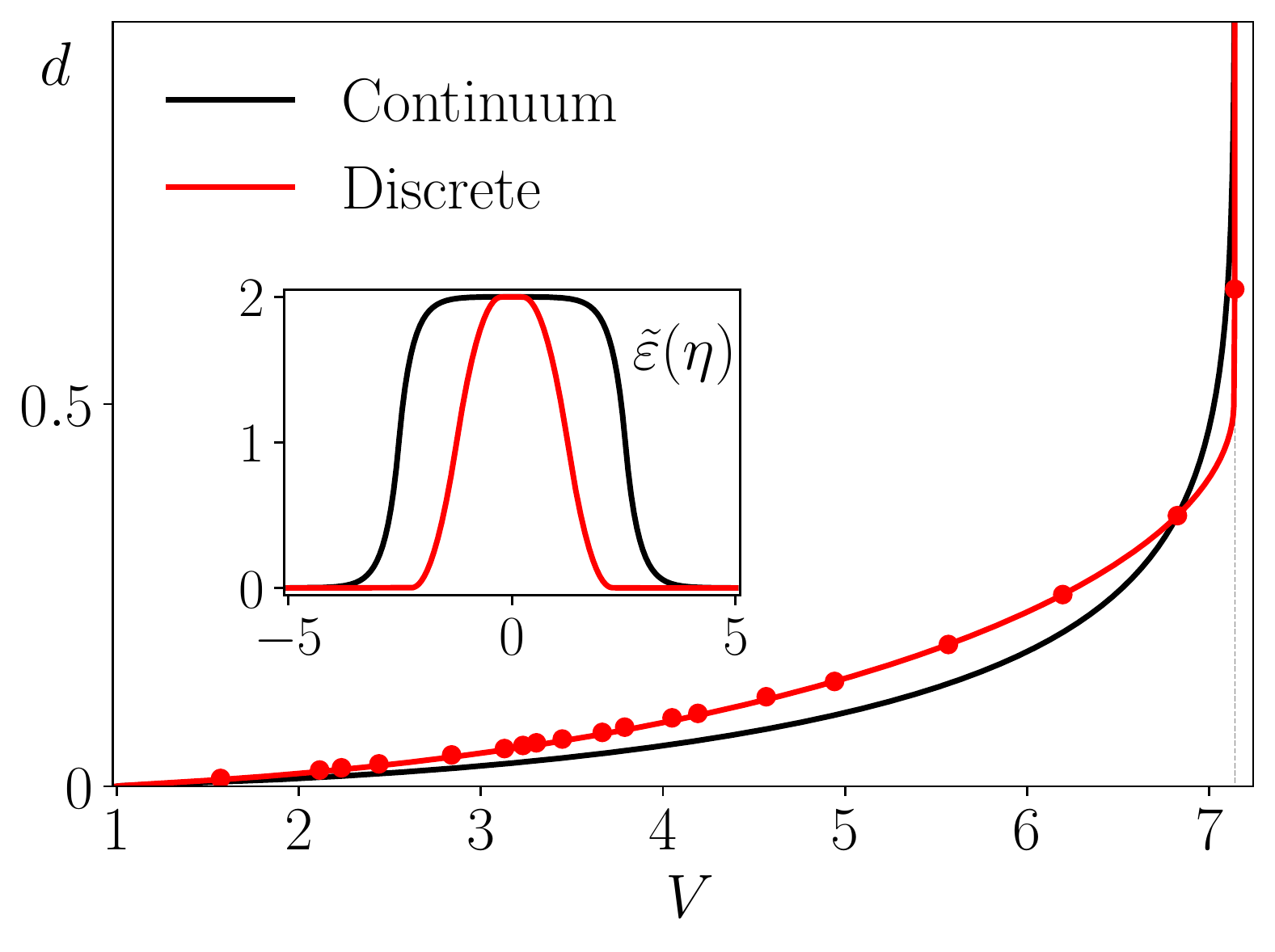}  
    \subfigimg[width=0.495\linewidth]{(b)}{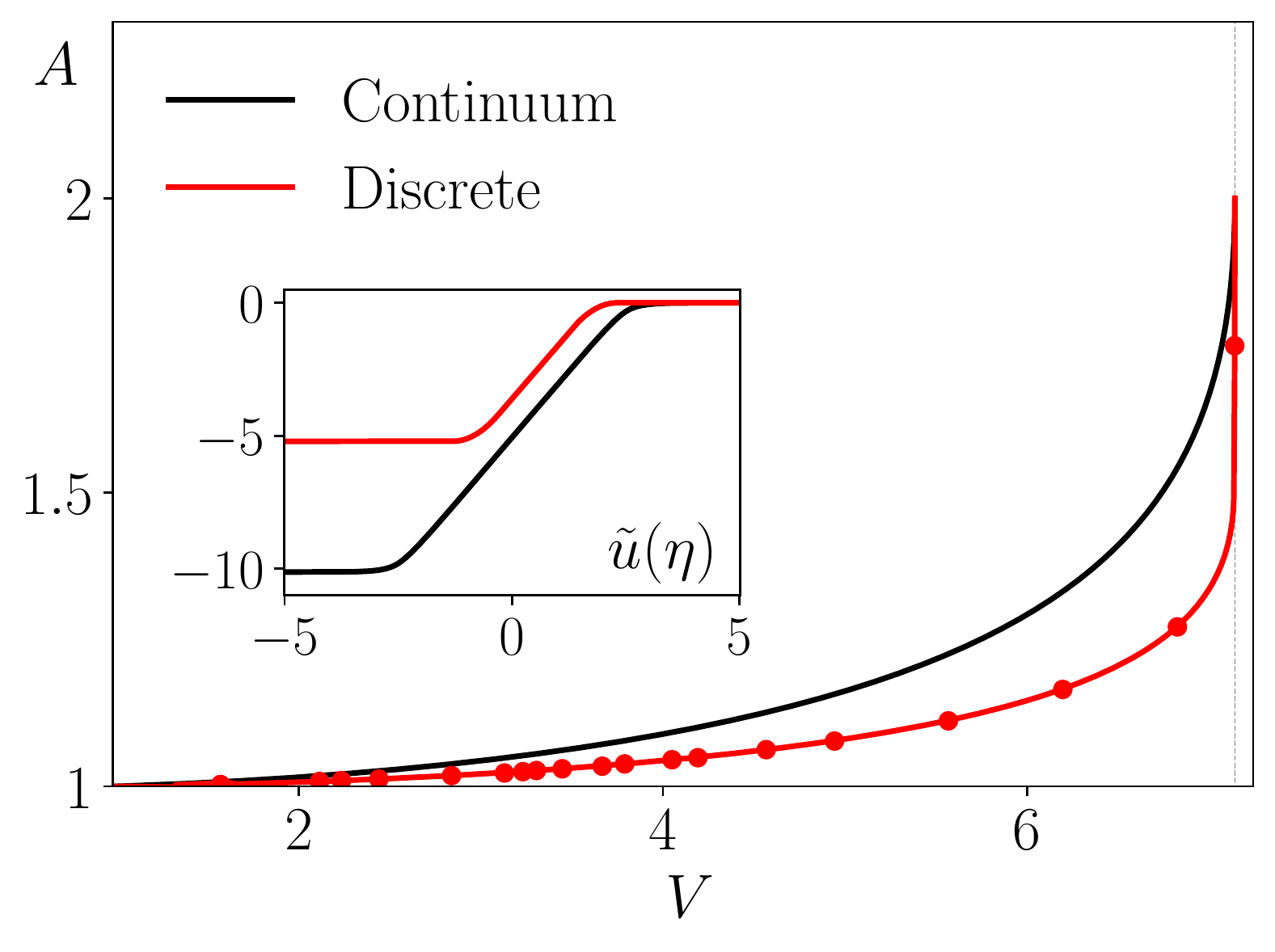}  
\caption{Comparison of  the predictions of discrete (red) and  continuum (black) models at $\tilde\sigma_a=100$ : (a)   half-width, (b)  amplitude. The red markers show the discrete pulses  obtained numerically by  solving  a one-parametric set of  initial value problems. The inserts compare discrete and continuum  distributions of strains  (a) and displacements (b) at the near critical speed $V=V_*-10^{-7}$.}
\label{fig:solution_discrete}
\end{figure} 
 
To construct solutions representing \emph{stretching}  pulses  we must solve \eqref{FPU} with the  stress   given by the same $\rect$ function as in the continuum case and use  the same matching condition  $\tilde\varepsilon(\pm d)=1$. The linearity of the  system at fixed $d$ hints that  the solitary wave  solution can be obtained  as a linear combination of two kinks
$
\tilde\varepsilon(\eta)=\tilde\varepsilon_k(\eta-d)-\tilde\varepsilon_k(\eta+d).
$
The nonlinear relation $d=d(V)$ can be then found from  the matching condition
$\tilde\varepsilon_k(0)-\tilde\varepsilon_k(2d)=1$. By inverting the  equation $(e^{-ik}-1)\hat{u}(k)=\hat{\varepsilon}(k)$ in the Fourier space we can also find the   displacement field $u(\eta)$, see~\cite{SOM}, and  compute the total   displacement which is again 
$ \Delta \tilde u= 2d\lambda. $ The typical dependencies $d(V),$ and $A(V)$ for \emph{stretching} pulses are illustrated  in Fig.~\ref{fig:solution_discrete}, where they are compared   with the corresponding  results of the continuum theory with $h=1$; the comparison confirms the overall adequacy of  the   continuum approximation  despite the choice of a finite value for  our  'small' parameter.

\begin{figure}[!h]
\begin{center}
\subfigimg[width=0.495\linewidth]{(a)}{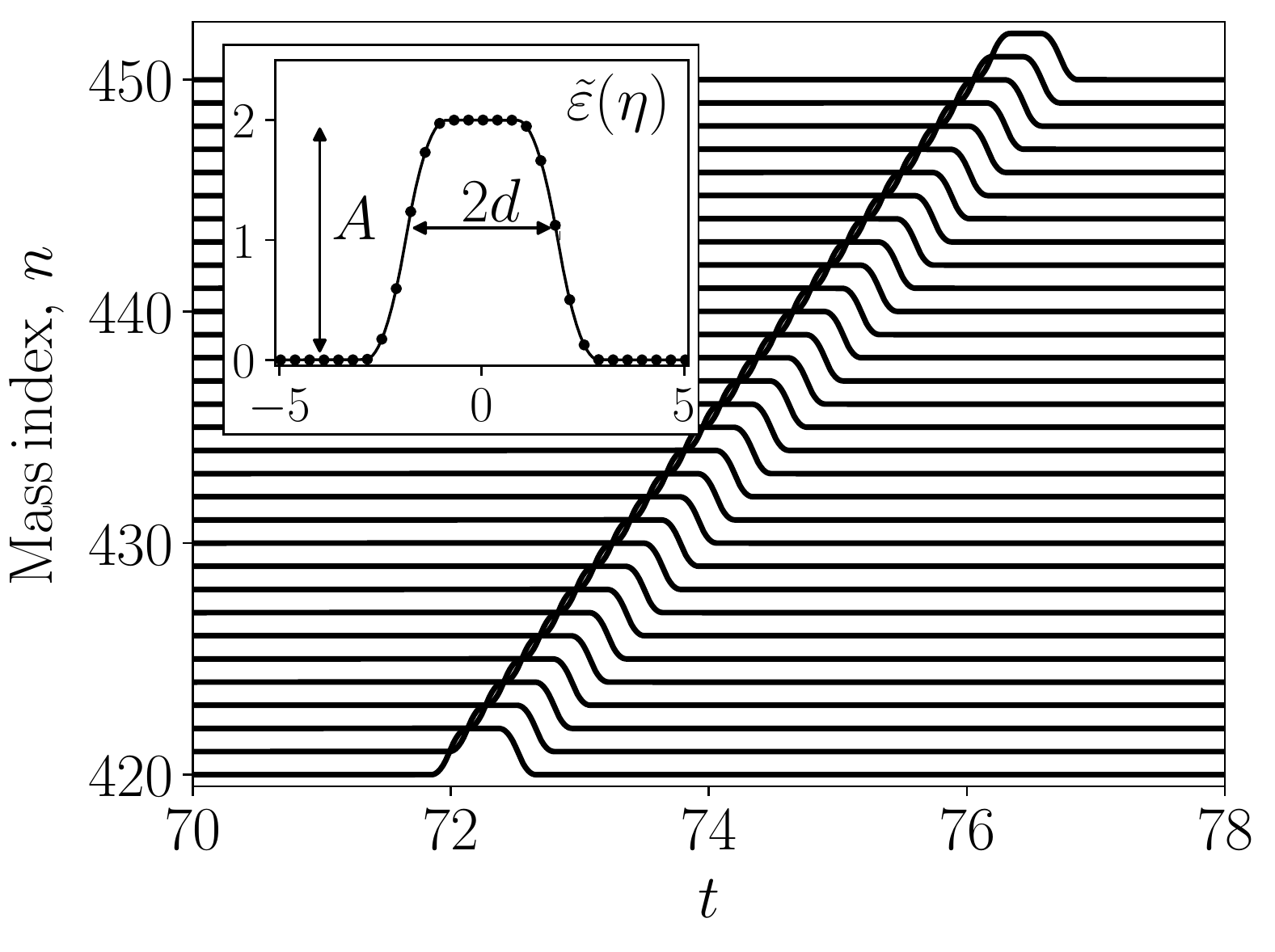} 
\subfigimg[width=0.495\linewidth]{(b)}{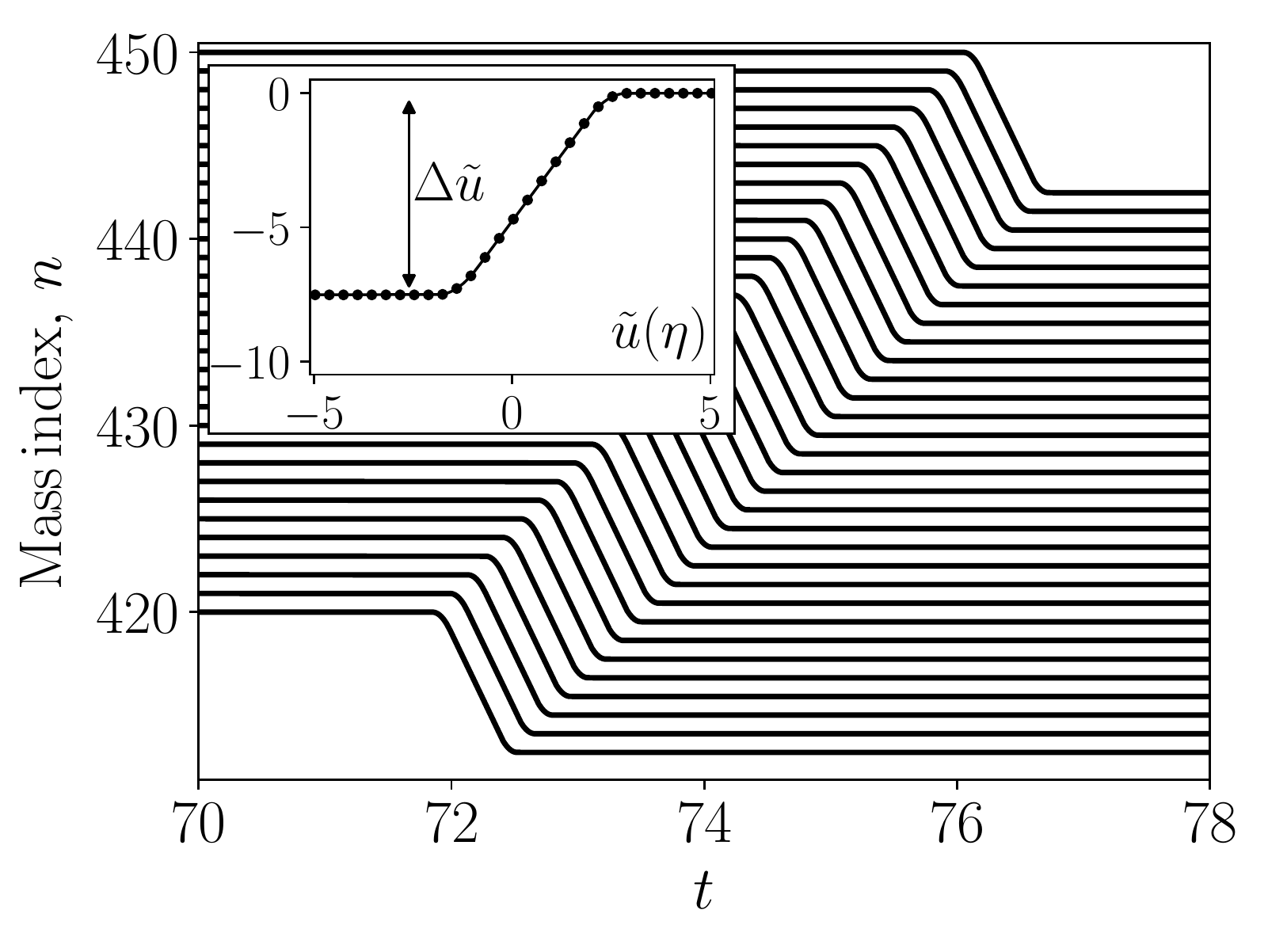}  
\end{center}
\caption{Results of numerical experiments  with discrete chain at $\tilde\sigma_a=100$ and  $v_0=200$ (see the text): (a) strains, (b) displacements.   The generated  stretching pulse  correspond to  $V\approx V_*$,  $d \approx 1.88$ and $A \approx 1.998$. Insets show the comparison of the numerical solution in the co-moving frame  (markers)   with the corresponding analytical solution (solid line). }
\label{fig:numerics}
\end{figure}
We  tested the  numerical stability  of the obtained discrete pulses   by  solving a range of   initial value problems.  An example of a stable  propagation of a rectangular stretching pulse is  shown in Fig.~\ref{fig:numerics}(a)  for the  chain with $1000$ springs. We  used initial conditions $\tilde u_n(0)=0,\,\dot{\tilde u}_1(0)=v_0$ and $\dot{\tilde u}_n(0)=0,\, n>1$ and assumed  free ends. The time evolution of the displacement field, shown  in Fig.~\ref{fig:numerics}(b),  illustrates the  creation of the main driver of peristaltic motility,  the finite displacement behind the propagating pulse.

To conclude, we developed a model of a dynamic passive-to-active  transformation  taking place  in  the front  of a steadily  moving   pulse with   the corresponding  reverse transformation  taking place  in its   rear. The resulting solitary wave solutions  for the active medium  were extended as  periodic  trains and  used to model  the peristaltic mode of self propulsion.  We found that at the critical value of parameter the model generates singular regimes  with diverging effective correlation length and argued that such criticality may be functional. Our results can be used for  biomimetic reproduction of worm-like motion   in  applications ranging from  endoscopic diagnostics~\cite{stefanini2006modeling} to pipeline inspection~\cite{yamashita2011self}. While in the existing robotic systems activity is usually imitated  by globally synchronized distributed actuators \cite{agostinelli2018peristaltic,miller2020gait,jiang2017optimal} our focus  on  local mechanical feedback in muscle-type  soft deformable  materials opens   new avenues in the  modeling of peristalsis, see also \cite{pehlevan2016integrative,umedachi2016gait,desimone2020cell}.
The proposed prototypical model  can serve only as a proof of concept and  future work   allowing one to make quantitative predictions, should  incorporate  energy supply and dissipation  and  account for  realistic 3D geometry.
 
The authors  thank G. Mishuris, P. Recho, A. Vainchtein for helpful discussions and  acknowledge the support of the French Agence Nationale de la Recherche  under the grant ANR-17-CE08-0047-02.


\end{document}


\title{Supplementary material for the paper: ``Peristalsis by  pulses of activity"}

\author{N.Gorbushin}
\affiliation{\it  PMMH, CNRS -- UMR 7636, CNRS, ESPCI Paris, PSL Research University, 10 rue Vauquelin, 75005 Paris, France}

\author{L. Truskinovsky}
\affiliation{\it  PMMH, CNRS -- UMR 7636, CNRS, ESPCI Paris, PSL Research University, 10 rue Vauquelin, 75005 Paris, France}

\date{\today}

\maketitle

\section{Quasi-continuum  problem}
The travelling waves in this model solve the  differential equation:
\begin{equation}
\left(V^2-\frac{V^2}{12}\frac{d^2}{d\eta^2}\right)\tilde\varepsilon(\eta)=\tilde\sigma(\eta)+\tilde\varepsilon_+,
\end{equation}
where $\tilde\varepsilon_+$ is  the integration constant defined by the condition that  $\tilde\varepsilon(\eta)\to \tilde\varepsilon_+$ when $\eta\to\infty$. In the case of stretching pulses, we look for solutions with $V>1$. The chosen ansatz for  $\tilde\sigma(\eta)$ distinguishes  between a kink, a solitary wave and  a  train of pulses which are  heteroclinic, homoclinic and periodic solutions, respectively. In what follows we adopt the normalization and the notations from the main text.

Kinks are characterized by a single transition event at $\eta=0$ and the corresponding equation reads
\begin{equation}
\left(V^2-\frac{V^2}{12}\frac{d^2}{d\eta^2}\right)\tilde\varepsilon(\eta)=\tilde\sigma(\eta)+\tilde\varepsilon_+,\quad \tilde\sigma(\eta)=\tilde\varepsilon(\eta)+\tilde\sigma_a H(-\eta).
\end{equation}
 The integration of this equation gives
\begin{equation}
\tilde{\varepsilon}_k(\eta)=\begin{cases}
\vspace{0.3cm}
\tilde{\varepsilon}_++\dfrac{\lambda}{2} e^{-\eta/z},\quad \eta>0,\\
\tilde{\varepsilon}_- -\dfrac{\lambda}{2} e^{\eta/z},\quad \eta<0,
\end{cases}
\label{eq:Solution_Kink_quasi}
\end{equation}
where   $\lambda=\tilde\sigma_a/(V^2-1)$ and $z=V/\sqrt{12(V^2-1)}$ and  $\tilde\varepsilon_-=\tilde\varepsilon_++\lambda$. Imposing the matching condition $\tilde\varepsilon(0)=1$ we obtain the closure condition $\tilde{\varepsilon}_\pm = 1 \mp \lambda/2$ which is equivalent to the requirement  that $V=V_*$  and that the two areas in Fig.1 of the main text are equal: $S_1=S_2$. Integration of the equation $\tilde \varepsilon_k(\eta)=d \tilde u_k(\eta)/d\eta$ allows one to reconstruct the  displacement field
\begin{equation}
\tilde{u}_k(\eta)=\begin{cases}
\vspace{0.2cm}
\tilde{\varepsilon}_+\eta -\dfrac{\lambda z}{2} e^{-\eta/z},\quad \eta>0,\\
\tilde{\varepsilon}_-\eta +\dfrac{\lambda z}{2} e^{\eta/z},\quad \eta<0,
\end{cases}
\end{equation}

 Solitary wave  solutions satisfy the condition  $\tilde\varepsilon_{\pm}=0$. The governing equation takes the form
\begin{equation}
\left(V^2-\frac{V^2}{12}\frac{d^2}{d\eta^2}\right)\tilde\varepsilon(\eta)=\tilde\sigma(\eta),\quad \tilde\sigma(\eta)=\tilde\varepsilon(\eta)+\tilde \sigma_a \rect{\left(\frac{\eta}{2d}\right)}.
\label{eq:Problem_Soliton_quasi}
\end{equation}
The active stress $\tilde \sigma_a$ is now applied on the finite interval $2d$ and we should  integrate the above equation under the assumption that  $\tilde{\varepsilon}(\pm d)=1$. 

Solution for this problem exist in the interval $1<V\leq V_{*}$. The    strain and displacement fields  can be written down explicitly
\begin{equation}
\tilde{\varepsilon}(\eta)=\begin{cases}
\vspace{0.2cm}
  e^{-(\eta-d)/z},\, \eta>d,\\
\vspace{0.2cm}
\lambda+(1-\lambda)\dfrac{\cosh(\eta/z)}{\cosh(d/z)},\, -d<\eta<d,\\
  e^{(\eta+d)/z},\, \eta<-d,
\end{cases}
\tilde{u}(\eta)=\begin{cases}
\vspace{0.2cm}
 -z e^{-(\eta-d)/z},\, \eta>d,\\
\vspace{0.2cm}
\lambda (\eta-d)+z(1-\lambda)\dfrac{\sinh(\eta/z)}{\cosh(d/z)},\, -d<\eta<d,\\
-\Delta \tilde u + z e^{(\eta+d)/z},\, \eta<-d.
\end{cases}
\end{equation}
To obtain the field  $\tilde u(\eta)$ we again set the trivial integration constant to 0 by assuming that  $\tilde u(\eta)\to0$ when $\eta\to\infty$. To close the system we must impose  the continuity of the derivative $d\tilde\varepsilon/d\eta$ at $\eta=\pm d$. We can  then compute the function  
\begin{equation}
d(V)=-z\tanh^{-1}\left(\frac{1}{1-\lambda}\right),
\label{eq:d_A_u_quasi}
\end{equation}
and obtain   other  important relations:
\begin{equation}
\quad A=\frac{1-\lambda}{\cosh(d/z)},\quad \Delta \tilde u=2d\lambda.
\label{eq:d_A_u_quasi1}
\end{equation}

The model describes  not only  the stretching  but also the contraction pulses. To obtain the latter we need to solve  the equation
\begin{equation}
\left(V^2-\frac{V^2}{12}\frac{d^2}{d\eta^2}\right)\tilde\varepsilon(\eta)=\tilde\sigma(\eta),\quad \tilde\sigma(\eta)=\tilde\varepsilon(\eta)+\tilde\sigma_a \left[1-\rect{\left(\frac{\eta}{2d}\right)}\right].
\end{equation}
 The boundary conditions must be now  chosen in the form  $\tilde\varepsilon\to \lambda$ when $\eta\to\pm\infty$. 
 
 Solution for this problem exist in the interval $V_*<V\leq V_{**}$. It can be again written explicitly
\begin{equation}
\tilde{\varepsilon}(\eta)=\begin{cases}
\vspace{0.2cm}
 \lambda+(1-\lambda)e^{-(\eta-d)/z},\, \eta>d,\\
\vspace{0.2cm}
\dfrac{\cosh(\eta/z)}{\cosh(d/z)},\, -d<\eta<d,\\
  \lambda+(1-\lambda)e^{(\eta+d)/z},\, \eta<-d,
\end{cases}
\tilde{u}(\eta)=\begin{cases}
\vspace{0.2cm}
\lambda(\eta-d)-z(1-\lambda) e^{-(\eta-d)/z},\, \eta>d,\\
\vspace{0.2cm}
z\dfrac{\sinh(\eta/z)}{\cosh(d/z)},\, -d<\eta<d,\\
\Delta \tilde u +\lambda(\eta-d)+ (1-\lambda)z e^{(\eta+d)/z},\, \eta<-d.
\end{cases}
\label{eq:Solution_contraction_quasi}
\end{equation}
The parameters here are
\begin{equation}
d=-z\tanh^{-1}(1-\lambda),\quad  A=\lambda - \frac{1}{\cosh(d/z)},\quad \Delta \tilde u=2d\lambda.
\end{equation}
This  solution degenerates in  two limiting cases. First,   at $V =1$ the integration  of \eqref{eq:Problem_Soliton_quasi} gives $\tilde{\varepsilon}(\eta)=6\tilde\sigma_a(d^2-\eta^2)+1$ when $-d<\eta<d$ and $\tilde\varepsilon(\eta)=1$ when $|\eta|>d$. Imposing the continuity of derivatives at  $\eta=\pm d$ we obtain  $d=0$ and $A=0$ and, therefore, $\tilde\varepsilon(\eta)\equiv 1$ and $\tilde \sigma(\eta)\equiv 1$.   In the other singular limit $V=V_{**}$ we have  $\lambda=1$ and therefore from  \eqref{eq:Solution_contraction_quasi} we find that $\tilde{\varepsilon}=1$ for $|\eta|>d$. The continuity of strain derivatives at $\eta=\pm d$ again gives $d=0$ and $A=0$, hence  $\tilde\varepsilon(\eta)\equiv 1$ but now $\tilde \sigma(\eta)\equiv 1+\tilde\sigma_a$.

To obtain the trains of active pulses we need to solve the equation 
\begin{equation}
\left(V^2-\frac{V^2}{12}\frac{d^2}{d\eta^2}\right)\tilde\varepsilon(\eta)=\tilde\sigma(\eta),\quad 
 \tilde\sigma(\eta)=\tilde\varepsilon(\eta)+\tilde\sigma_a\sum_{j}\rect\left(\frac{\eta_j}{2d_p}\right),
\end{equation}
where $\eta_j=\eta-2jD$. The integration of this equation gives with the matching conditions mentioned in the main text gives
\begin{equation}
\tilde{\varepsilon}(\eta)=
\begin{cases}
\vspace{0.2cm}
\dfrac{e^{d_p/z}}{1+e^{2(d_p-D)/z}}\left(e^{(\eta_j-2D)/z}+e^{-\eta_j/z}\right),\quad d_p<\eta_j\leq D,\\
\vspace{0.2cm}
\lambda+(1-\lambda)\dfrac{\cosh(\eta_j/z)}{\cosh(d_p/z)},\quad -d_p<\eta_j<d_p,\\
\dfrac{e^{d_p/z}}{1+e^{2(d_p-D)/z}}\left(e^{\eta_j/z}+e^{-(\eta_j+2D)/z}\right),\quad -D< \eta_j< -d_p.
\end{cases}
\label{eq:Soliton_train_quasi}
\end{equation}
The value of the parameter $d_p$ can be found   as a positive real root of the transcendental equation
\begin{equation}
\left(1-\lambda\right)\tanh(d_p/z)=\tanh((d_p-D)/z),
\label{eq:d_train_quasi}
\end{equation}
which is equivalent to the requirement of continuity for the  first derivative of strain  at $\eta_j=\pm d_p$. 
By integrating \eqref{eq:Soliton_train_quasi}, while respecting the continuity of displacements, we obtain 
\begin{equation}
\tilde{u}(\eta)=
\begin{cases}
\vspace{0.2cm}
j\Delta\tilde u+\dfrac{z e^{d_p/z}}{1+e^{2(d_p-D)/z}}\left(e^{(\eta_j-2D)/z}-e^{-\eta_j/z}\right),\quad d_p<\eta\leq D,\\
\vspace{0.2cm}
j\Delta\tilde u+\lambda \left(\eta_j-d_p\right)+z\left(1-\lambda\right)\dfrac{\sinh(\eta_j/z)}{\cosh(d_p/z)},\quad -d_p<\eta_j< d_p,\\
(j-1)\Delta\tilde u+\dfrac{ze^{d_p/z}}{1+e^{2(d_p-D)/z}}\left(e^{\eta_j/z}-e^{-(\eta_j+2D)/z}\right),\quad -D<\eta_j<-d_p.
\end{cases}
\end{equation}

\section{Discrete problem}
The  discrete problem reduces to a solution of the advance-delay differential equation
\begin{equation}
V^2\frac{d^2\tilde\varepsilon}{d\eta^2}=\tilde\sigma(\eta-1)+\tilde\sigma(\eta+1)-2\tilde\sigma(\eta).
\label{eq:Problem_discrete}
\end{equation}
 In view of the partial  linearity of the problem, it will be  enough to construct the kink-type solution while individual pulses and  trains of pulses can be obtained  as linear combinations of kinks with appropriately adjusted continuity conditions. As above, the kink solution can be obtained if use in  \eqref{eq:Problem_discrete} the ansatz  $\tilde\sigma(\eta)=\tilde\varepsilon(\eta)+\tilde\sigma_a H(-\eta)$. We can then apply the Fourier transform $\hat{\varepsilon}_k(k)=\int_{-\infty}^{\infty}\tilde\varepsilon_k(\eta)e^{ik\eta}\,dk$ and rewrite  the problem as an algebraic one
\begin{equation}
L(k)\hat{\varepsilon}_k(k)=-\tilde\sigma_a\frac{\omega^2(k)}{0+ik},
\end{equation}
where the kernel is $L(k)=\omega^2(k)-(kV)^2$, $\omega^2(k)=4\sin^2(k/2),$ is the dispersion relation and $(0+ik)^{-1}=\lim_{\alpha\to0+}(\alpha+ik)^{-1}$ stands for the Fourier transform of the Heaviside function . 

The solution of the original problem can be  presented in the form of the inverse Fourier transform
\begin{equation}
\tilde{\varepsilon}_k(\eta)=\tilde{\varepsilon}_+-\frac{\tilde\sigma_a}{2\pi}\int_{-\infty}^{\infty}\frac{\omega^2(k)}{(0+ik)L(k)}e^{-ik\eta}\,dk,
\end{equation}
where the contour integrals can be computed by the residue method.   The kernel function $L(k)$ has  a double zero at $k=0$. The rest of the roots   are simple and complex,  located in  both half-planes. They can be organized in   the sets $Z^\pm=\left\{k\,:\,L(k)=0,\,\pm \mathrm{Im}\, k>0\right\}$. Then the explicit series solution of the discrete problem can be written in the form:
\begin{equation}
\tilde{\varepsilon}_k(\eta)=\begin{cases}
\vspace{0.2cm}
\tilde{\varepsilon}_++\tilde\sigma_a\sum\limits_{k\in Z^-}\dfrac{\omega^2(k)}{kL'(k)}e^{-ik\eta},\, \eta>0,\\
\tilde{\varepsilon}_- -\tilde\sigma_a\sum\limits_{k\in Z^+}\dfrac{\omega^2(k)}{kL'(k)}e^{-ik\eta},\, \eta<0.
\end{cases}
\label{eq:Solution_kink}
\end{equation}
Here $\tilde{\varepsilon}_+$ is a homogeneous solution of the problem with the boundary conditions  $\tilde\varepsilon(\eta)\to\tilde\varepsilon_{\pm}$ at $\eta \to \pm \infty$ and where $\tilde{\varepsilon}_-=\tilde{\varepsilon}_++\lambda$. 

To apply the matching condition at $\eta=0$ we need to consider  an infinitely large  circle in the complex plane.  The contour integration in this case gives the relation
\begin{equation}
\lambda+\tilde\sigma_a\sum_{k\in Z^-}\dfrac{\omega^2(k)}{kL'(k)}+\tilde\sigma_a\sum_{k\in Z^+}\dfrac{\omega^2(k)}{kL'(k)}=0.
\end{equation}
If we also recall that   $L(-k)=L(k)$ and $L(\overline{k})=\overline{L(k)}$ we obtain $\sum_{k\in Z^-}\omega^2(k)/(kL'(k))=\sum_{k\in Z^+}\omega^2(k)/(kL'(k))$ and, hence, $\tilde\varepsilon(0)=\tilde\varepsilon_+-\lambda/2=\tilde\varepsilon_-+\lambda/2$. By applying the matching condition  $\tilde{\varepsilon}_\pm=1\mp\lambda$ we obtain again the equal area construction $S_1=S_2$, illustrated in 
 Fig.1(a) of the main text.  The obtained condition means that the entropy production on the corresponding jump discontinuity in the coarse grained continuum  problem would be equal to  zero.

\begin{figure}[!h]
  \centering
  \includegraphics[scale=0.3]{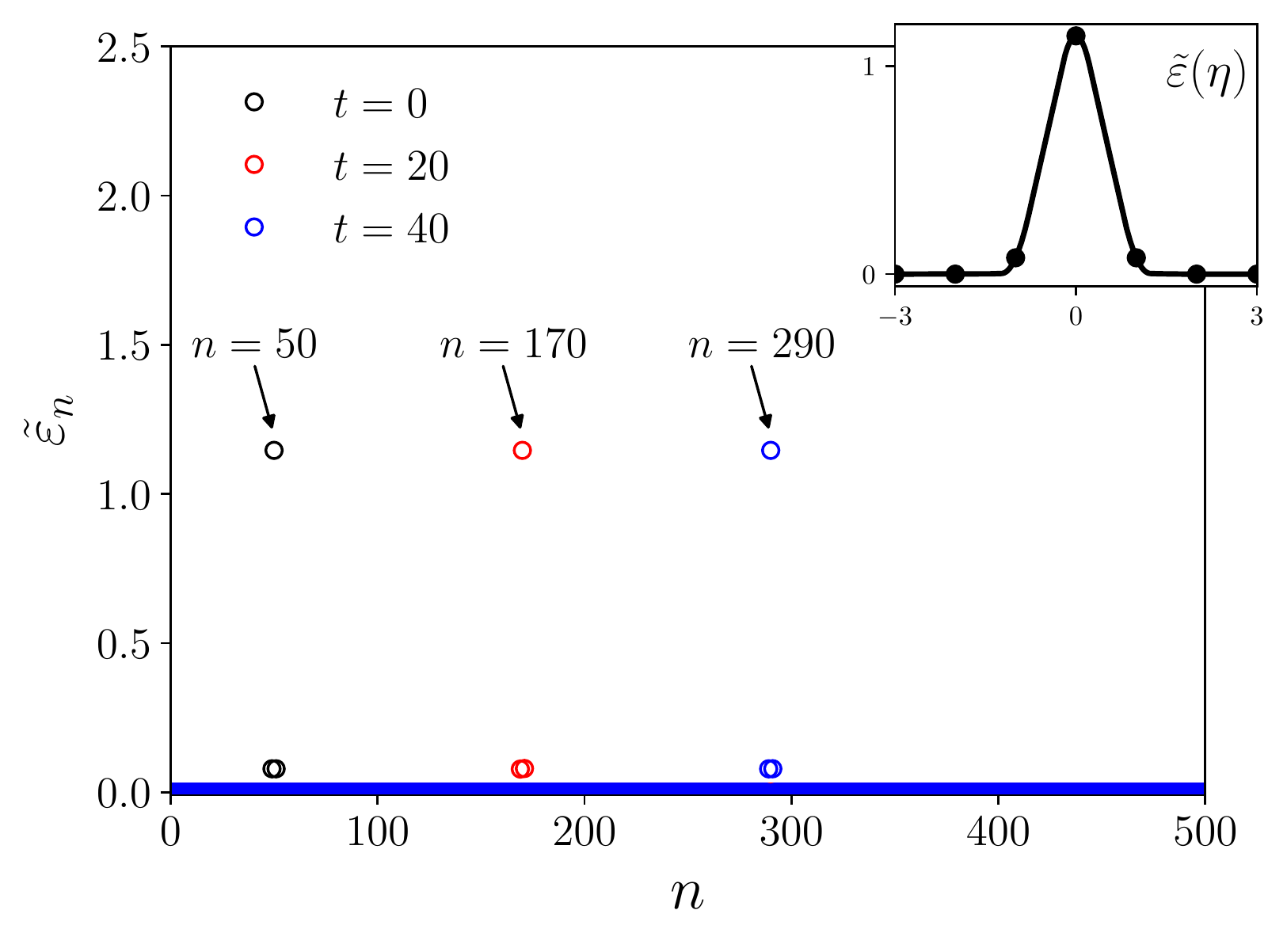}
  \caption{Stable propagation of a stretching pulse  at $V=6$ and $\tilde\sigma_a=100$. The inset shows the superimposed analytical solution.}
\label{fig:soliton_num}
\end{figure}

We can also construct the discrete    displacement by inverting the    relation  $(e^{-ik}-1)\hat{u}_k(k)=\hat{\varepsilon}_k(k)$. Following the same scheme as above we obtain  
\begin{equation}
\tilde{u}_k(\eta)=\begin{cases}
\vspace{0.2cm}
\tilde{\varepsilon}_+(\eta-1/2)+\tilde\sigma_a\sum_{k\in Z^-}\dfrac{i\omega^2(k)}{k\sin(k/2)L'(k)}e^{-ik(\eta-1/2)},\, \eta>1/2,\\
\tilde{\varepsilon}_-(\eta-1/2)-\tilde\sigma_a\sum_{k\in Z^+}\dfrac{i\omega^2(k)}{k\sin(k/2)L'(k)}e^{-ik(\eta-1/2)},\, \eta<1/2.
\end{cases}
\label{eq:Solution_kink_displacements}
\end{equation}
Using  \eqref{eq:Solution_kink} and \eqref{eq:Solution_kink_displacements} we can construct   the solution for the  stretching pulse  using the ansatz   $\tilde\sigma(\eta)=\tilde\varepsilon(\eta)+\tilde\sigma_a\rect(\eta/(2d))$. It takes the form  $\tilde\varepsilon(\eta)=\tilde\varepsilon_k(\eta-d)-\tilde\varepsilon_k(\eta+d)$ for the discrete strain field and $\tilde u(\eta)=\tilde u_k(\eta-d)-\tilde u_k(\eta+d)$ for the discrete displacement field, where $d$ is found from the equation $\tilde{\varepsilon}_k(0)-\tilde{\varepsilon}_k(2d)=1$.  The stable propagation of a discrete pulse  is illustrated in  Fig.~\ref{fig:soliton_num} for a discrete chain with $500$ springs.

For the contraction pulse we need to use another ansatz  $\tilde\sigma(\eta)=\tilde\varepsilon(\eta)+\tilde\sigma_a[1-\rect(\eta/(2d))]$. The discrete strain field can be then written in the form $\tilde\varepsilon(\eta)=\tilde\varepsilon_k(d+\eta)+\tilde\varepsilon_k(d-\eta)$.  For the   displacement field we obtain  $\tilde u(\eta)=\tilde u_k(d+\eta)+\tilde u_k(d-\eta)$. The value of  the parameter  $d$ in the corresponding interval of velocities $V_*<V\leq V_{**}$ can be found from the equation $\tilde{\varepsilon}_k(0)+\tilde{\varepsilon}_k(2d)=1$. 

Using the same idea we can   find the solution describing the  train of   stretching pulses. In this case we need to use  the ansatz  $\tilde\sigma(\eta)=\tilde\varepsilon(\eta)+\sum_j\tilde\sigma_a\rect(\eta_j/(2d_p))$. The strain and displacement fields  are now  represented  via infinite sums $\tilde{\varepsilon}(\eta)=\sum_j \tilde{\varepsilon}_k(\eta_j-d)-\tilde{\varepsilon}_k(\eta_j+d_p)$ and $\tilde{u}(\eta)=\sum_j\tilde{u}_k(\eta_j-d)-\tilde{u}_k(\eta_j+d_p)$. The equation for finding $d_p$ ( see the main text) is now 
\begin{equation}
\sum_{j}\tilde{\varepsilon}_k(-2jD)-\tilde{\varepsilon}_k(2d_p-2jD)=1.
\end{equation}
In the discrete case we can also define  the parameter  $d=\min{(d_p,D_p-d_p)}$  observing  that  $d_p=D/2$ at $V=V_*$.